\documentclass[useAMS,usenatbib]{mn2e}
\usepackage{epsfig}
\usepackage{float}
\usepackage[T1]{fontenc}
\usepackage{graphicx}
\usepackage{epstopdf}

\usepackage{natbib}

\usepackage{color}

\usepackage{aas_macros}

\usepackage{array}
\newcolumntype{L}[1]{>{\raggedright\let\newline\\
\arraybackslash\hspace{0pt}}m{#1}}
\newcolumntype{C}[1]{>{\centering\let\newline\\
\arraybackslash\hspace{0pt}}m{#1}}
\newcolumntype{R}[1]{>{\raggedleft\let\newline\\
\arraybackslash\hspace{0pt}}m{#1}}

\def\sige{\mbox{$\sigma_{\rm e}$}}

\def\Re{\mbox{$R_{\rm e}$}}

\def\Msun{\mbox{$M_\odot$}}

\def\ML{\mbox{$M/L$}}
\def\dimf{\mbox{$\delta_{\rm IMF}$}}

\def\Yst{\mbox{$\Upsilon_\star$}}

\def\mst{\mbox{$M_{\star}$}}

\def\Mvir{\mbox{$M_{\rm vir}$}}
\def\cvir{\mbox{$c_{\rm vir}$}}

\def\lsim{\mathrel{\rlap{\lower3.5pt\hbox{\hskip0.5pt$\sim$}}
    \raise0.5pt\hbox{$<$}}}                
\def\gsim{~\rlap{$>$}{\lower 1.0ex\hbox{$\sim$}}}
\def\Rap{\mbox{$R_{\rm Ap}$}}
\def\sigAp{\mbox{$\sigma_{\rm Ap}$}}

\def\Fig{\mbox{Fig.~}}

\def\Sec{\mbox{Sect.~}}

\def\Eq{\mbox{Eq.~}}

\def\a0{\mbox{$a_{\rm 0}$}}

\def\gN{\mbox{$g_{\rm N}$}}

\title[ETGs and emergent gravity]{Testing Verlinde's emergent gravity in early-type galaxies}

\author[Tortora C. et al.]{\noindent
C.~Tortora$^{1}$\thanks{E-mail: ctortora@astro.rug.nl},
L.~V.~E.~Koopmans$^{1}$, N.~R.~Napolitano$^{2}$,
E.~A.~Valentijn$^{1}$
\\~\\
$^1$ Kapteyn Astronomical Institute, University of Groningen, P.O.
Box 800, 9700 AV Groningen, the Netherlands \\
$^2$ INAF -- Osservatorio Astronomico di Capodimonte, Salita
Moiariello, 16, 80131 - Napoli, Italy\\
}

\begin{document}
\date{Accepted  Received }
\pagerange{\pageref{firstpage}--\pageref{lastpage}} \pubyear{xxxx}
\maketitle

\label{firstpage}
\begin{abstract}
Verlinde derived gravity as an emergent force from the information
flow, through two-dimensional surfaces and recently, by a priori
postulating the entanglement of information in three-dimensional
space, he derived the effect of the gravitational potential from
dark matter (DM) as the entropy displacement of dark energy by
baryonic matter. In Emergent Gravity (EG) this apparent DM depends
only on the baryonic mass distribution and the present-day value
of the Hubble parameter. In this paper we test the EG proposition,
formalized by Verlinde for a spherical and isolated mass
distribution, using the central dynamics (SDSS velocity
dispersion, $\sigma$) and the K-band light distribution in a
sample of $4260$ massive ($\mst \gsim 10^{10} \, \rm \Msun$) and
local early-type galaxies (ETGs) from the SPIDER datasample. Our
results remain unaltered if we consider the sample of 807 roundest
field galaxies. Using these observations we derive the predictions
by EG for the stellar mass-to-light ratio (\ML) and the Initial
Mass Function (IMF), and compare them with the same inferences
derived from a) the standard DM-based models, b) an alternative
description of the missing mass (i.e. MOND) and c) stellar
population models. We demonstrate that, consistently with a
classical Newtonian framework with a DM halo component, or
alternative theories of gravity as MOND, the central dynamics can
be fitted if the IMF is assumed non-universal. The results can be
interpreted with a IMF lighter than a standard Chabrier at
low-$\sigma$, and bottom-heavier IMFs at larger $\sigma$. We find
lower, but still acceptable, stellar \ML\ in EG theory, if
compared with the DM-based NFW model and with MOND. The results
from EG are comparable to what is found if the DM haloes are
adiabatically contracted and also with expectations from spectral
gravity-sensitive features. If the strain caused by the entropy
displacement would be not maximal, as adopted in the current
formulation, then the dynamics of ETGs could be reproduced with
larger \ML.
\end{abstract}


\section{Introduction}\label{sec:intro}

Dark matter (DM) is one of the biggest puzzles in the modern
astrophysics and cosmology. This unseen mass component is thought
to dominate the mass density of galaxies and clusters of galaxies
in the Universe. It is essential to explain the high orbital
velocities of gas and stars in the outer regions of spiral
galaxies (\citealt{Bosma78_PhD}) and leaves its imprint at
cosmological scales (e.g., \citealt{Komatsu+11_WMAP7}). DM is
elusive, interacting very weakly with visible matter and has not
yet detected by any experiment. Thus, alternative ways to solve
the missing mass problem have been suggested (e.g. MOND,
\citealt{Milgrom83,Milgrom83b}), since it could be related to our
poor understanding of gravity at the galactic and cluster scales.
However, all kinds of approaches to solve the missing mass problem
need to properly account for the stellar and gas content in
galaxies and clusters. The first step in relating DM to visible
matter of stars in the centers of galaxies (we neglect the
contribution of gas in this paper) is the appreciation of the
effective overall stellar mass-to-light ratio (\ML) of stellar
populations, which today is still considered to be
ill-constrained. In particular, the uncertainties on the number of
low-mass stars can induce a change in the \ML\ of stars of about a
factor 2 (e.g. \citealt{Tortora+09}). The distribution of stars in
a galaxy, and thus the ratio between low- and high-mass stars, is
described by the so called stellar Initial Mass Function (IMF).
Most of the studies of resolved stellar populations are obviously
only possible in the Milky Way, where IMF was originally
characterized as a power-law mass-distribution, $dN/dM \propto
M^{-\alpha}$, with $\alpha \sim 2.35$ \citep{Salpeter55}, and
subsequently refined to flatten at lower masses ($M \lsim 0.5
M_\odot$; \citealt{Kroupa01,Chabrier03}). Chabrier- or Kroupa-like
IMFs have been usually adopted in most types of galaxies,
environments and redshifts. Recently, this hypothesis has been
questioned by different lines of observational evidence
(\citealt{vDC10}; \citealt{Treu+10}; \citealt{ThomasJ+11};
\citealt{Conroy_vanDokkum12b}; \citealt{Cappellari+12,
Cappellari+13_ATLAS3D_XX}; \citealt{Spiniello+12};
\citealt{Wegner+12}; \citealt{Barnabe+13}; \citealt{Dutton+13};
\citealt{Ferreras+13}; \citealt{Goudfrooij_Kruijssen13};
\citealt{LaBarbera+13_SPIDERVIII_IMF}; \citealt{TRN13_SPIDER_IMF};
\citealt{Weidner+13_giant_ell}; \citealt{Goudfrooij_Kruijssen14};
\citealt{Shu+15_SLACSXII}; \citealt{McDermid+14_IMF};
\citealt{Tortora+14_DMslope,Tortora+14_MOND};
\citealt{Martin-Navarro+15_IMF_variation};
\citealt{Spiniello+15_IMF_vs_density}; \citealt{Lyubenova+16};
\citealt{TLBN16_IMF_dwarfs}; \citealt{Corsini+17};
\citealt{Li+17_IMF}; \citealt{Sonnenfeld+17_IMF}). In the current
analysis we will express our results in terms of stellar \ML\ and
the associated IMF, discussing whether these are realistic within
different gravity frameworks.

As mentioned before, one of the alternatives to DM is the MOdified
Newtonian Dynamics (MOND) by \cite{Milgrom83b, Milgrom83}, which
proposes that the missing mass problem in galaxies could be
resolved by a modification of Newton's law in the extremely
low-acceleration regime. Newton's second law of dynamics becomes
$F = m g$, where the acceleration $g$ is related to the Newtonian
one \gN\ by $g \, \mu(g/\a0) = \gN$. Here, \a0~$\sim cH_0$ (where
c is the speed of light and $H_0$ the Hubble constant) and
$\mu(x)$ is the {\it interpolating function}, with the limiting
behaviours $\mu(x\gg1) = 1$ and $\mu(x\ll1) = x$. It has been
shown to reproduce flat rotation curves of spiral galaxies without
the need for DM, and naturally predicting the observed relation
between galaxy rotation and luminosity
\citep{TF77,Sanders_McGaugh02} or baryonic mass
(\citealt{McGaugh12}). Only a few MOND analyses have been carried
out on early-type galaxies (ETGs; e.g., \citealt{Cardone+11MOND};
\citealt{Ferreras+12_MOND_TEVES}; \citealt{Milgrom+12_Xray_ell};
\citealt{Sanders14}). Recently, \cite{Tortora+14_MOND} have
demonstrated that MOND is consistent with the central dynamics of
ETGs if the stellar \ML\ are systematically larger in
higher-velocity dispersion ($\sigma$) galaxies, when compared with
those calculated from stellar populations, assuming a universal
stellar IMF. Thus, in MOND, the IMF is non universal, suggesting
that it is ``lighter'' for low-$\sigma$ galaxies, and ``heavier''
for high velocity dispersions. This result is consistent with a
plethora of independent studies using central dynamics (and/or
gravitational lensing) with standard DM halo models or modelling
spectral gravity-sensitive features (see above for a list of
references).

A completely different proposal to supersede the presence of DM in
the most massive virialized structures has been recently suggested
(\citealt{Verlinde11,Verlinde16}). This idea proposes that
space-time and gravity are macroscopic concepts that arise from an
underlying microscopic description in which these concepts have no
meaning. We refer to this proposition as Emergent Gravity (EG) and
use the formalism in \cite{Verlinde16}. EG modifies gravity at
scales set by the `Hubble acceleration scale' $a_0$. Similarly to
MOND, the gravitational force emerging in the EG framework exceeds
that of General Relativity (GR) on supergalactic and cluster
scales. However, the underlying physical modeling in EG is rather
different. Unlike MOND and DM-based models, the (apparent) DM
distribution only depends on the baryonic mass distribution
$M_b(r)$ and $H_0$. But, most important, it depends on the radial
mass density gradients, which could provide the ultimate tool for
testing and comparing the validity of EG. Note that the formalism
for EG has currently only been derived for spherical symmetric
baryonic mass distributions, and any test should account for these
still very first and simplified predictions.

Some papers have recently tested this new proposition on different
scales, using weak gravitational lensing signal
(\citealt{Brouwer+16_EG}), dynamics of dwarf spheroidal satellites
of the Milky Way (\citealt{Diez-Tejedor+16_EG_dSph}), dwarf
galaxies (\citealt{Pardo17_EG}), the radial acceleration relation
in spiral galaxies (\citealt{Lelli+17_Verlinde}), galaxy rotation
curves and Solar System planets' perihelia (\citealt{Hees+17_EG}),
X-ray galaxy clusters (\citealt{Ettori+16_Verlinde}), lensing and
X-ray in clusters (\citealt{Nieuwenhuizen+16}), finding
contrasting results. \cite{Milgrom_Sanders16_EG} have highlighted
some limits in the Verlinde's EG proposition.

In this paper, we will test the EG predictions in the central
regions of massive (with stellar masses $\mst \gsim 10^{10} \, \rm
\Msun$) and local (with redshifts $0.05 < z < 0.095$) ETGs from
the SPIDER survey (\citealt{SPIDER-I}). We fix the Hubble
parameter and allow the only free parameter of the model, i.e.,
the stellar \ML, to vary, evaluating our results in terms of the
IMF. Then, we compare the EG results with MOND, DM haloes and with
findings from gravity sensitive spectral features. We will also
analyze how the Hubble constant value, and thus the acceleration
scale, impact our results, discussing how the strain caused by the
entropy displacement can be constrained. The paper is organized as
follows. In \Sec\ref{sec:EG} we introduce the EG proposition,
discussing hypothesis and approximations. Standard models for DM
halos and MOND theory are described in
\Sec\ref{sec:Newton_and_MOND}. The data sample is presented in
\Sec\ref{sec:sample_SPIDER}. In \Sec\ref{sec:results} we will
discuss the results of our dynamical analysis, and finally
\Sec\ref{sec:conclusions} is devoted to discussion and
conclusions.

\section{Emergent gravity}\label{sec:EG}

Emergent Gravity (EG) refers to the idea that space-time and
gravity are macroscopic effects arising from underlying
microscopic physical phenomena. In many fields of the physics,
macroscopic phenomena "emerge" from microscopic processes, for
example thermodynamics arising from microscopic states of matter,
or the Van der Waals force emerging from non-relativistic quantum
electrodynamics. Examples are found in different physical
environments, and thus it is not inconceivable that the nature of
gravity could be the macroscopic and cumulative effects of some
unknown microscopic physics. Moreover, this would not be
surprising considering that classical General Relativity already
has signatures which link it to thermodynamics: e.g., black hole
thermodynamics. These ideas are not new, and different approaches
have been followed to describe the emergent gravity (e.g.
\citealt{Sindoni12_EG}).

In the context of EG, \cite{Jacobson95} has shown how Einstein
equations can be recovered from the black hole entropy and the
standard concepts of heat, entropy, and temperature in
thermodynamics. The idea that gravity is an entropic force was
further explored by \cite{Padmanabhan10} and \cite{Verlinde11}.
They suggested that an "area law" scaling of gravitational
entropy, as opposed to the usual volume scaling, is essential to
derive Einstein's laws of gravity. Motivated by all these ideas,
\cite{Verlinde16} has used these thermodynamic concepts to suggest
a possible alternative explanation for DM. He suggests that the
quasi-de Sitter spacetime, which would be a fair approximation of
the present Universe, dominated by dark energy, can be obtained
from a system of microstates which are coherently excited above
the true vacuum. This ground state corresponds to an anti-de
Sitter spacetime filled by a negative cosmological constant, and
emerges from microstates which are entangled. Thus,
\cite{Verlinde16} uses the theory of elasticity
(\citealt{Padmanabhan04_elasticity_gravity}) and has suggested
that in addition to the area law, a volume entanglement of entropy
can be postulated. The stress between the area law in
\cite{Verlinde11} and the volume law in \cite{Verlinde16}
manifests itself in spherical galaxies  as an apparent DM on
scales set by the 'Hubble acceleration scale' $a_{0} = c H_{0}$,
where c is the speed of light and $H_0$ the present-day Hubble
parameter. By studying the dynamics of galaxies and clusters, it
is possible to test the evidence for this postulate.

Thus, according to \cite{Verlinde16}, there exists an extra
gravitational effect, in addition to the classical GR contribution
of $M_{\rm b}(r)$ to the gravitational potential. This excess in
the gravity emerges due to the volume law contribution to the
entropy, that is associated with the dark energy content of our
universe. In a universe without matter, the total entropy of the
dark energy would be maximal, as it would be homogeneously
distributed over the microstates. In our Universe, on the other
hand, its baryonic mass distribution $M_{\rm b}(r)$ reduces the
entropy content of the universe. This removal of entropy due to
matter produces an elastic response of the underlying microscopic
system, which can be observed on galactic and supergalactic scales
as an additional gravitational force. The difference with GR is
that this excess gravity does not come from the existence of DM.
However, the excess mass in the Verlinde's proposal can still be
described by an {\it apparent} DM distribution $M_{\rm DM}(r)$.

\subsection{Formulation}

We start with some qualitative arguments. A system with a static,
spherically symmetric and isolated baryonic mass distribution is
analyzed (see \citealt{Verlinde16} for further details). We
consider a spherical region with boundary area $A(r) = 4\pi
r^{2}$, which contains a mass $M$ near its center. The surface
mass density $\Sigma_{\rm M}(r)$ is defined as the ratio of the
mass M and the area $A(r)$. Empirically, the directly observed
gravitational phenomena attributed to DM is thought to occur when
the surface mass density falls below a universal value determined
by the acceleration scale \a0\ (e.g. \citealt{Milgrom83b}). This
condition could be written as $\Sigma_{\rm M}(r) < \a0 / (8\pi
G)$, where $\a0 = c H_{\rm 0}$. This inequality is made more clear
if written in terms of de Sitter entropy removed by adding the
mass M, i.e. $S_{\rm M}$, and the one related to dark energy,
$S_{\rm DE}$. In this regime, we assume the inequality $S_{\rm M}
< S_{\rm DE}$ holds. The nature of gravity changes depending on
whether matter removes all or just a fraction of the de Sitter
entropy. In general, we can define the strain $\epsilon_{\rm M}(r)
\equiv S_{\rm M}/S_{\rm DE} = 8\pi G \Sigma{\rm M} / \a0$. If
$\epsilon_{\rm M}(r) > 1$ the dynamics of stellar objects behaves
as in Newtonian framework, while if $\epsilon_{\rm M}(r) < 1$,
then we are in the regime of low surface mass density and low
acceleration, i.e. in the "dark gravity" regime. $\epsilon_{\rm
M}$ corresponds to the largest principle of the elastic medium
strain. Thus, when only a part of the de Sitter entropy is removed
by matter inclusion, the remaining entropy induces a non
negligible effect, leading to modifications of the normal
gravitational laws in the Newtonian regime. This translates into
an "apparent" surface density produced by baryons, $\Sigma_{DM}=
(\a0\epsilon_{\rm DM})/(8\pi G)$. To determine these
modifications, we would need to analyze the displacement of the
entropy content, due to matter, applying the linear elasticity
theory (\citealt{Verlinde16}).

The quantity of apparent DM can be obtained by estimating the
elastic energy associated with the entropy displacement caused by
$M_{\rm b}(r)$. After some calculations, this leads to the
following relation\footnote{We avoid to report the lengthy
calculations made in \cite{Verlinde16}. The reader can refer to
that paper for further information.}:
\begin{equation}
    \int_0^r \varepsilon_{\rm DM}^2 (r') A(r')dr' \leq V_{M_{\rm b}} (r)  \, ,
    \label{eq:fundamentalrelation}
\end{equation}
where we integrate over the sphere with radius $r$ and area
$A(r)$. The strain $\varepsilon_{\rm DM}(r)$ caused by the entropy
displacement is given, as defined previously, by:
\begin{equation}
    \varepsilon_{\rm DM} (r) = \frac{8\pi G}{c H_0} \frac{M_{\rm DM}(r)}{A(r)} \, ,
    \label{eq:strain}
\end{equation}
Furthermore, $V_{M_{\rm b}}(r)$ is the volume that would contain
the amount of entropy that is removed by a mass $M_{\rm b}$ inside
a sphere of radius $r$, if that volume were filled with the
average entropy density of the universe:
\begin{equation}
    V_{M_{\rm b}}(r) = \frac{8\pi G}{c H_0} \frac{M_{\rm b} (r) \, r}{3} \, .
    \label{eq:removedvolume}
\end{equation}
\Eq\ref{eq:fundamentalrelation} deserves some attention, because
due to the inequality, observations can only put a lower bound on
the $M_{\rm DM}$ and $H_{\rm 0}$, since a larger value can be
accommodated by having a smaller elongation (or compression) of
the elastic medium due to the baryonic mass inclusion. Throughout
this paper, we assume that the largest principle strain
$\varepsilon_{\rm DM} (r)$ takes its maximal value and the
response of the medium is negligible outside the mass. These
assumptions authorize us to adopt the equality in
\Eq\ref{eq:fundamentalrelation}.

Thus, inserting the relations (\ref{eq:strain}) and
(\ref{eq:removedvolume}) into (\ref{eq:fundamentalrelation}), and
taking the derivative with respect to $r$ on both sides of the
equation, one arrives at the following relation:
\begin{equation}
    M_{\rm DM} (r) = r \sqrt{\frac{cH_0}{6G}}  \sqrt{\frac{d\left( M_{\rm b}(r) r \right)}{dr}} \, .
    \label{eq:veg_mdm}
\end{equation}
This is the apparent DM formula, which translates a baryonic mass
distribution into an apparent DM distribution. As it emerges from
\Eq\ref{eq:veg_mdm}, the EG formalism naturally provides the value
of a characteristic acceleration scale, $a_{EG} \equiv c H_{\rm
0}/6$.

In our analyses, the mass from \Eq\ref{eq:veg_mdm} is added to
baryons and the resulting mass is converted into a velocity
dispersion, using the standard Jeans equations and treating the
apparent DM in EG as a real mass component. These predicted
velocity dispersions are subsequently compared with the observed
velocity dispersions. We will discuss this procedure in more
details in the following sections.

\subsection{Caveats and assumptions}\label{subsec:caveats}

Currently \Eq\ref{eq:veg_mdm} is the only specific prediction of
the DM in the EG framework and we will use it in the rest of this
paper, recognizing the following caveats and restrictions, which
as a matter of fact also apply to various other recent papers on
EG predictions (\citealt{Brouwer+16_EG};
\citealt{Diez-Tejedor+16_EG_dSph}; \citealt{Ettori+16_Verlinde};
\citealt{Nieuwenhuizen+16}; \citealt{Lelli+17_Verlinde};
\citealt{Hees+17_EG}; \citealt{Pardo17_EG}). In particular, we
will discuss: a) the assumption of spherical symmetry and
isolation of the mass distribution
(\Sec\ref{subsubsec:spherical_isolated}), b) the cosmological
framework, motivating our assumptions about the present-day Hubble
parameter (\Sec\ref{subsubsec:cosmology}), c) the limitations of
the equality in \Eq\ref{eq:fundamentalrelation}
(\Sec\ref{subsubsec:lower_bound}), and finally d) our assumption
that the Jeans equations are the same as used in a Newtonian or
MONDian framework (\Sec\ref{subsubsec:Jeans}).

\subsubsection{ETGs: spherically symmetric and isolated}\label{subsubsec:spherical_isolated}

The predictions of the excess mass in EG is currently only valid
for static, spherically symmetric and isolated baryonic mass
distributions. Therefore, one of the best test-bench for
Verlinde's EG is represented by early-type galaxies, which match
these characteristics. As we will discuss, massive ETGs are among
the best galaxy candidates, since many of them are approximatively
spherically symmetric. ETGs contain most of the stellar mass of
the universe, and represent the final stage of galaxy evolution.
In a standard $\Lambda$CDM scenario, they are thought to be fossil
records of the stellar and DM assembly through time. In this
scenario, DM is dominant in the external regions, while the
extremely complex physics of gas and stars are dominant in the
central regions. ETGs exhibit a peaked surface brightness profile
and historically have been considered to be well fitted by a
\cite{deVauc48} profile, in contrast to late-type systems which
present more extended and shallow light distributions, which are
described by exponential profiles. However, more detailed analyses
show that their light distributions are well described by the
S\'ersic law (\citealt{Ciotti91}), with a shape parameter, $n$
(S\'ersic index), that accounts for variations of the light
profile shape among galaxies ($n=4$ corresponds to a
\citealt{deVauc48} profile). These steep profiles are also
accompanied by the absence of disk and spiral arm structures, and
are thought to be the result of accretion (\citealt{Hilz+13};
\citealt{Tortora+14_DMevol}).

ETG shapes are well described by an oblate or triaxial ellipsoid,
and when projected on the sky, they have small ellipticities,
becoming rounder toward larger masses
(\citealt{Vulcani+11_ellipticity}). However, also when a residual
small ellipticity is present, as in most ETGs, the spherical
approximation is far better than a similar assumption for spiral
galaxies, which are characterized by a central bulge and a
pronounced disk, which is far to be approximated by a sphere. The
difference between spherical and disk geometry can induce
corrections of the order of $\sim 20\%$ for spirals
(\citealt{Lelli+17_Verlinde}).

Moreover, ETGs are found to live in all galaxy environments, from
the field to groups and clusters of galaxies. From this point of
view they are also good candidates to test the Verlinde's
proposal, which has produced an expression for the excess mass for
an object only if it is sufficiently far from any other mass
distribution and unaffected by recent or ongoing merging-events or
close interactions.

\subsubsection{Cosmological framework}\label{subsubsec:cosmology}

In order to test EG predictions, we need to make some assumptions
about the adopted cosmology, which enters in the distances and the
evolution of the Hubble parameter in \Eq\ref{eq:veg_mdm}.
Verlinde's arguments only hold in a static Newtonian
approximation, allowing one to describe gravity phenomena on
galactic and super-galactic scales, but they are not sufficient to
include the evolution of the Universe. His EG proposal has been
only developed in a de Sitter space-time, which relies on the
approximation that our universe is entirely dominated by dark
energy (in particular by a cosmological constant $\Lambda$) and
that standard baryonic matter only leads to a small perturbation.
Two main issues arise from these assumptions. First, in a de
Sitter space-time, the Hubble parameter can be written as $H(z) =
\mathcal{H}_{0} \sqrt{\Omega_{\Lambda}} \propto \sqrt{\Lambda}$,
which means is constant with time. This motivates the assumption
about $H_{0} = H(z=0) = H(z)$ adopted by Verlinde in
\Eq\ref{eq:veg_mdm}. Another approximation is in the assumption
that the dark energy is the dominant contributor to the energy
density of the Universe. This is an incorrect approximation mainly
at the early stages of the Universe, when the contribution of the
dark energy is smaller compared with other energy contributors.

In a standard $\Lambda$CDM framework, the following formula holds
$H(z)=\mathcal{H}_{0}\sqrt{\Omega_{\rm
m}(1+z)^{3}+\Omega_{\Lambda}}$, and from the comparison with CMB
spectrum it is found that the Universe is flat and thus
$H(z=0)=\mathcal{H}_{0}$ (e.g., \citealt{Komatsu+11_WMAP7}). But
this is obviously not true in a de Sitter space-time, where the
Hubble parameter at $z=0$ depends on the cosmological constant.

A de Sitter space-time is not a realistic model capable to fit the
cosmological data, and in particular cannot fit the $H(z)$
inferred from standard cosmological data (e.g. redshift-distance
relation in Type-Ia Supernoavae, \citealt{Riess+98_SN}; CMB
anisotropies, \citealt{Komatsu+11_WMAP7}; the baryon acoustic
oscillation, BAO, peaks, \citealt{Percival+10_BAO}), nor
observational $H(z)$ data using passive galaxies as 'cosmic
chronometers' (OHD, \citealt{Jimenez+03_OHD};
\citealt{Zhang+14_OHD}). However, at low redshifts, the shape of
$H(z)$ determined from observational probes are almost independent
on the exact cosmological model adopted. This is true since
various models are fitting the data producing local $H(z)$ values
which are consistent with $H_{0}$ from $\Lambda$CDM cosmology
within the measurement uncertainties (e.g.,
\citealt{Carvalho+08_fR}; \citealt{Zhang+14_OHD};
\citealt{Farooq+17}). For the redshifts studied in this paper,
i.e. $z < 0.1$, the cosmological evolution has a negligible effect
on the distance measurements and on Verlinde's equations. Thus, we
assume that observations at $z \sim 0$ can be reproduced by an
effective $\Lambda$CDM model with $\Omega_{m}=0.3$,
$\Omega_{\Lambda}=0.7$ and $H_0= 75 \, {\rm km \, s^{-1}
Mpc^{-1}}$. Here we stay in line with all of the recent papers
that have directly or indirectly adopted a background $\Lambda$CDM
cosmology (\citealt{Brouwer+16_EG};
\citealt{Diez-Tejedor+16_EG_dSph}; \citealt{Ettori+16_Verlinde};
\citealt{Lelli+17_Verlinde}; \citealt{Hees+17_EG}) to set the
value of $H_{0}$ in \Eq\ref{eq:veg_mdm} and the distances, and
some of them investigated the impact of a varying $a_{0}$. Again
in line with these papers, we will also discuss the case when
$a_{0}$, and thus $H_{0}$, changes (e.g.
\citealt{Lelli+17_Verlinde}; \citealt{Hees+17_EG}).

\subsubsection{Only a lower bound on DM distribution}\label{subsubsec:lower_bound}

Verlinde, as in all the recent literature which addressed his
proposal, have assumed that the excess gravity generated from the
elastic response and the related apparent DM distribution take
their maximal value. This translates to an equality in
\Eq\ref{eq:fundamentalrelation}. If we consider the inequality in
the Verlinde's formula, then to a fixed amount of baryons will
correspond a smaller $M_{\rm DM}$. The strain would assume
reasonably the largest value sufficiently close to the bulk of
mass distribution, where the contribution of the apparent DM first
becomes noticeable (\citealt{Verlinde16}). But, getting further
out, or when other mass distributions come in, the inequality
hold. Thus, in the central regions of ETGs this assumption would
seem more reasonable, but it does remain arbitrary and needs to be
tested. Thus, the relaxation of the equality in our assumption
would provide some constraints on the entropy strain.

\subsubsection{Jeans equations}\label{subsubsec:Jeans}

We will interpret  measured velocity dispersions in terms of
gravity by applying the Jeans theorem, and assume this is
justified in an EG formalism. In this section we will motivate why
this is the case. ETGs are self-gravitating systems of stars with
random motions, which can be quantified by the velocity
dispersion. In a standard Newtonian framework, the motions of
stars in the gravitational potential are described by the Jeans
equations, which are derived from the collisionless Boltzmann
equation and relate the components of the velocity dispersion of
the system (e.g., in polar coordinates, $\sigma_{r}$,
$\sigma_{\theta}$ and $\sigma_{\phi}$) to the gravitational
potential $\phi(r)$, and thus to the mass distribution
(\citealt{Binney_Tremaine87}). The radial Jeans equation is the
relevant formula in this paper, and gets a very simple expression
if one assumes a) a steady-state hydrodynamic equilibrium, b)
spherical symmetry, and c) a single tangential velocity dispersion
$\sigma_t$:
\begin{equation}
{{\rm d}[j_{\rm \star}(r) \sigma_{r}^{2}(r)] \over {\rm d}r} +
2\,{\beta(r) \over r} \,j_{\rm \star}(r) \sigma_{r}^2(r) = -
j_{\rm \star}(r)\, \frac{GM(r)}{r^2} \ , \label{eq:jeans_N}
\end{equation}
where $j_{\rm \star}(r)$ and $M(r)$ are the deprojected light and
total mass distribution, and $\beta = 1 - \sigma_t^2/\sigma_r^2$
is the anisotropy. This equation holds in the Newtonian case, but
can be easily generalized to account for generic acceleration
function $g(r)$, and in particular for MOND formalism,
as\footnote{We notice that there is a typo on the right side of
Equation 2 in \cite{Tortora+14_MOND}.}
\begin{equation}
\frac{d[j_{\rm \star}(r) \sigma_{r}^{2}(r)]}{dr} +2 \frac{
\beta(r)}{r} j_{\rm \star}(r) \sigma_{r}^{2}(r)=-j_{\rm \star}(r)
g (r), \label{eq:jeans_g}
\end{equation}
which assumes the standard expression in \Eq\ref{eq:jeans_N} when
is $g(r) = G M(r)/r^{2}$.

We can use one of the previous two formulas, by just replacing
$M_{DM}(r) + \mst(r)$ in \Eq\ref{eq:jeans_N}, or, using the MOND
formalism, evaluating the centripetal acceleration associated to
this apparent total mass in \Eq\ref{eq:jeans_g}, consistently with
the interpolating function provided in \cite{Hees+17_EG}. The
radial velocity dispersion from the Jeans equations has to be
first integrated along the line of sight and then projected within
a finite aperture (rectangular or circular; \citealt{Tortora+09}).
Then, this quantity can be compared to the observed aperture
velocity dispersion, \sigAp, to determine the stellar \ML\ and/or
the present-day Hubble parameter, the only free parameters in
\Eq\ref{eq:veg_mdm}.

We do not find any strong argument against a similar formalism in
the Verlinde's framework, where the hypothesis and approximations
discussed to obtain the radial Jeans equations seem to be
satisfied. Thus, following \cite{Diez-Tejedor+16_EG_dSph} we adopt
the radial Jeans equation in \Eq\ref{eq:jeans_N}, inserting the
total mass, obtained summing up the baryonic and apparent DM mass
derived from \Eq\ref{eq:veg_mdm}, i.e. $M(r) = M_{\star}(r) +
M_{\rm DM}(r)$, assuming for the stellar distribution a constant
\ML\ profile. However, this assumption will need to be better
analyzed in the future.

\section{Newtonian and MONDian framework}\label{sec:Newton_and_MOND}

Here we will adopt some specific DM halo models and also MOND,
determining their best-fitted stellar \ML s and comparing these
results with those from EG. In the Newtonian framework, we adopt a
two-component model (stars + DM). For the DM distribution we
assume some DM halo models, fixing their parameters according to
viable recipes, which we will discuss below. While, for the EG and
MOND propositions, we use a constant \ML\ profile, with free
stellar \ML, \Yst, for the stellar distribution.

Results using these DM-based models have been presented in
\cite{TRN13_SPIDER_IMF} and \cite{Tortora+14_DMslope}. New results
using MOND with SPIDER dataset are presented here for the first
time. In \cite{Tortora+14_MOND} we tested MOND using a different
datasample, but finding similar results. In the rest of this
section, we provide details about all these models.

\subsection{DM-based models}\label{subsec:DM_models}

The DM profile from N-body simulations is well described by a
double power-law, commonly referred to as the NFW profile,
parameterized by two parameters, the virial concentration index
\cvir\ and the (total) virial mass \Mvir\ (\citealt{NFW96,
NFW97}). We adopt NFW as the reference DM model, assuming a) the
correlation between \Mvir\ and \cvir, from N-body simulations
based on WMAP5 cosmology \citep{Maccio+08}, as well as b) the
\Mvir--$\mst$ correlation from abundance matching results in
\citet{Moster+10}, assuming a Chabrier IMF for \mst. In this way,
for each galaxy with a stellar mass \mst, \Mvir\ and \cvir\ are
empirically set, and the DM profile is fully determined.

In order to explore the effect of a possible modification to the
DM profile because of the interaction between gas and stars with
DM, we also consider the case of an NFW profile with an adjustable
degree of baryon-induced adiabatic contraction (AC, e.g.,
\citealt{Blumenthal+86}; \citealt{Gnedin+04}). AC is an
approximate way to model the expected drag of dissipatively
infalling stars on the surrounding DM particles, producing a halo
with a higher central DM density than in collisionless N-body
simulations. Following \cite{NRT10} we adopt the \cite{Gnedin+04}
prescription.

We have also explored how these results depend on the assumed
\Mvir -- \cvir\ relation
(\citealt{TRN13_SPIDER_IMF,Tortora+14_DMslope}).

We have analyzed the impact of the mass density law adopting a
\cite{Burkert95} profile, which is the prototype of cored models,
and has been shown to reproduce the DM profile of spirals and
dwarf galaxies. The density and scale parameter of the Burkert
profile ($\rho_{\rm B}$ and $r_{\rm B}$, respectively) are assumed
to follow the relation from \cite{Salucci_Burkert00}, adjusted to
match results at higher surface density, for two ETGs, by
\citet{Memola+11}. We have explored two cases in detail, where the
scale radius is set to $r_{\rm B} = 1$ and $20 \, \rm kpc$,
respectively. We have shown in \cite{Tortora+14_DMslope} that the
exact value of $r_{\rm B}$ has a negligible impact on the inferred
stellar \ML\ values.

\subsection{MOND-based models}\label{subsec:MOND_models}

MOND assumes that standard dynamics is not valid in the limit of
low accelerations, such that the gravitational acceleration $g(r)$
differs from the Newtonian one $\gN(r) = G \mst/r^{2}$. The
MONDian $g(r)$ reduces to the Newtonian one at high accelerations.
In the low-acceleration limit, i.e. deep in the MONDian regime,
the acceleration is given by $(g/\a0)g = \gN$, where \a0\ is the
MOND acceleration constant. MOND predicts flat rotation curves in
the external regions of spiral galaxies and naturally leads to the
\cite{TF77} relation. The characteristic acceleration scale \a0\
is a fundamental parameter of the theory (\citealt{Milgrom83b}).
In this paper, we adopt the standard value of $\a0 =
1.2\times10^{-10}$~m~s$^{-2}$, as calibrated from spiral galaxy
dynamics (\citealt{Begeman+91}). This value is found to be of the
same order of magnitude as the "acceleration" associated to the
Hubble constant, i.e. $\approx c H_{\rm 0}$, and to the
cosmological constant $\Lambda$, $\approx c (\Lambda/3)^{1/2}$
(\citealt{Milgrom+01_MOND_review}). If we use the definition
provided in EG, i.e. $\a0 = a_{\rm EG} \equiv c H_{\rm 0} / 6$,
then the value adopted for \a0\ corresponds to $H_{\rm 0} \approx
75 \, \rm km s^{-1} Mpc^{-1}$.

To connect the low- and high-acceleration regimes, the following
expression is adopted:
\begin{equation}
g(r) \mu \left[ \frac{g(r)}{\a0} \right] = \gN(r) ,
\label{eq:MOND}
\end{equation}
where $x=g(r)/\a0$ and $\mu(x)$ is an empirical ``interploating''
function, with the properties $\mu(x\gg1) = 1$ and $\mu(x\ll1) =
x$. One recovers the Newtonian theory when $\mu(x)=1$ and the deep
MOND regime when $\mu(x)=x$. An alternative expression can be
obtained making the substitution $\nu(y) = \mu(x)^{-1}$, where
$y\equiv \gN/\a0$.

We adopt the following expressions:
\begin{itemize}
\item the first attempts to fit rotation curves adopted the interpolating function $\mu(x)= x/\sqrt{1+x^{2}}$
(\citealt{Milgrom83b}; \citealt{Sanders_McGaugh02});
\item later on another law has been suggested to provide a better description of some data,
i.e. the "simple" function $\mu(x)= x/(1+x)$
(\citealt{Famaey_Binney05}; \citealt{Angus08});
\item recently, using more than 2500 data points in a sample of 153 rotationally supported galaxies,  \cite{McGaugh+16}
suggested the following expression $\nu(y) = (1-
\exp(-\sqrt{y}))^{-1}$.
\end{itemize}
A constant \ML\ profile with a free \Yst\ is adopted for the total
mass distribution (see \citealt{Tortora+14_MOND} for further
details). To reduce the computation time, we follow the same
binning procedure used in \cite{TRN13_SPIDER_IMF}, constructing
``average'' galaxies by dividing our sample into different
\sige-bins, for which we compute median values of all the stellar
parameters $(\Re, n, \mst, \Rap, \sigAp)$. For each \sige-bin and
a given interpolation function, we solve the radial Jeans equation
\Eq\ref{eq:jeans_g} for the \Yst\ value matching the observed
average \sigAp\ in the bin.

\section{Datasample}\label{sec:sample_SPIDER}

In this section, we will describe the sample of galaxies used and
their main properties, that we use to test the DM, MOND and EG
models.

\subsection{SPIDER sample}

The SPIDER survey has demonstrated to be very useful in the study
of the luminous and DM distribution in the galaxy cores
(\citealt{SPIDER-VI, TRN13_SPIDER_IMF,Tortora+14_DMslope}). It
consists of a sample of $5,080$ bright ($M_r<-20$) ETGs, in the
redshift range of $z=0.05$ to $0.095$, with optical and
Near-InfraRed (NIR) photometry available ($grizYJHK$ bands) from
the Sloan Digital Sky Survey (SDSS) DR6 and the UKIRT Infrared
Deep Sky Survey-Large Area Survey
DR3~\footnote{http://www.sdss.org, http://www.ukidss.org}
(\citealt{SPIDER-I}). S\'ersic profile is fitted to the surface
photometry using 2DPHOT (\citealt{LaBarbera_08_2DPHOT}). Thus, the
effective radius \Re\ and S\'ersic index $n$ have been measured
from $g$ through $K$ bands. SPIDER ETGs have central velocity
dispersions, \sigAp, measured in the circular aperture of the SDSS
fiber ($\Rap = 1.5$ arcsec). The median ratio of the SDSS fibre to
the K-band effective radius is $\Rap/\Re \sim 0.6$. The \sige\ is
the SDSS-fibre velocity dispersion, $\rm \sigAp$, corrected to an
aperture of one \Re , following \cite{Cappellari+06}.

ETGs are defined  as  bulge-dominated systems  (i.e. SDSS
parameter $fracDev_r \!  > \!  0.8$, which measures the fraction
of galaxy light better fitted by a de~Vaucouleurs, rather than an
exponential law), with passive spectra within the SDSS fibres
(SDSS attribute $eClass \!   < \!  0$, where $eClass$ indicates
the spectral  type  of  a  galaxy  based on  a principal component
analysis). See \cite{SPIDER-I} for further details. For the
present work, we rely on a subsample of 4260 SPIDER ETGs, with
higher quality optical and NIR structural parameters, selected as
in \citet{SPIDER-VI}, with S\'ersic fits having $\chi^2<2$ in all
wavebands and uncertainty on $\log \Re$ $<$ $0.5$~dex, as well as
available stellar masses. For each galaxy, the stellar
population-based mass-to-light ratio, $\Yst_{\rm Chab}$, has been
determined by fitting \citet{BC03} stellar population models to
the multi-band photometry, under the assumption of a Chabrier IMF
(\citealt{SPIDER-V}; \citealt{SPIDER-VI}).

All these  galaxies reside on the red-sequence, with more than $99
\%$ having $g-r \gsim 0.5$, within an aperture of 1~\Re, and a
median $g-r=0.88$.  Stellar mass (Chabrier IMF-based) and aperture
velocity dispersions for the sample are in the ranges $\sim (0.1 -
3 \times 10^{11} \, \rm \Msun$ and $\sim 50-250\, \rm km s^{-1}$,
with medians of $5.4 \times 10^{10}\, \Msun$ and $154\, \rm km
s^{-1}$, respectively. The \Re\ and S\'ersic index values span the
ranges $\sim (0.5 - 40) \, \rm kpc$ and $\sim 2-10$, with medians
of $3.3 \, \rm kpc$ and 6.6.

\subsection{A test-bench for EG}

ETGs are the best candidates to test the Verlinde's model, since
they are the objects which approach the approximations made by
Verlinde (spherical symmetry and isolation) and can be found in
large numbers in local environments and higher redshift.

ETGs can have a wide range of shapes and in particular, the axis
ratios of the SPIDER galaxies have $q \gsim 0.2$, with a
distribution which is peaked at $q \sim 0.75$, with a median of
0.69. If we limit the analysis to the roundest galaxies, e.g.
imposing $q > 0.6$, then 2847 out of 4260 are left and the median
axis ratio is $q=0.77$. The environment of ETGs in the SPIDER
sample is characterized by a friends-of-friends catalog of $8083$
groups (\citealt{Berlind+06}; \citealt{Lopes+09}), classifying
galaxies as either group members, field galaxies, or unclassified.
We select the sample of 1230 field galaxies. See \cite{SPIDER-III}
for further details. The galaxy sample is left with 807 objects,
after both the criteria are applied.

In \cite{SPIDER-VI} we have found that the impact of the galaxy
ellipticity and environment on the DM fractions is negligible. A
similar result is found in \cite{TRN13_SPIDER_IMF}, fixing the DM
profile as discussed in \Sec\ref{subsec:DM_models}, and finding
that IMF is only negligible affected. These results further
support our choice of retaining the whole sample in our EG
analysis. We will also demonstrate that the results for EG remain
unchanged if only isolated and rounder galaxies are considered.

\section{Analysis and results}\label{sec:results}

We derive the dynamical (i.e. total) mass distribution of ETGs by
solving the spherical isotropic Jeans equations for the three
cases of EG, standard DM models and MOND (see
\ref{subsec:caveats}). A given model for the mass profile is
fitted to \sigAp\ for SPIDER\footnote{We use the dynamical
procedure described in \cite{Tortora+09} and
\cite{Tortora+14_DMslope}, no seeing correction is adopted.}. In a
Newtonian framework, two-component mass models, describing baryons
and DM, are adopted. With MOND, a model for baryons is adopted and
equations are modified to account for the change of the force law,
as discussed in \Sec\ref{subsec:caveats}
(\citealt{Tortora+14_MOND}).

We assume that gas contributes negligibly to the mass profile,
i.e. $M_{b} (r) = \mst(r)$ (\citealt{Courteau+14_review};
\citealt{Li+17_IMF}). Thus, the baryons are made up by only stars,
whose surface brightness is modeled by a S\'ersic profile. The
shape parameter $n$ and effective radius of the S\'ersic laws are
those obtained by fitting galaxy images in K-band (see
Sec.~\ref{sec:sample_SPIDER}). The light distribution is converted
into stellar mass by means of a constant stellar mass-to-light
ratio, \Yst, which is a free fitting parameter. We assume
negligible gradients in stellar populations
(\citealt{Tortora+11MtoLgrad}) and IMF
(\citealt{Martin-Navarro+15_IMF_variation}; \citealt{Alton+17}).

In the following section we discuss the results of the paper. We
will first set constraints on the present-day Hubble parameter in
\Sec\ref{subsec:results_H0}, then in \Sec\ref{subsec:results_IMF}
we fix $H_{\rm 0}$ and leave the stellar \ML\ free to change,
investigating the shape of the IMF, and finally in
\Sec\ref{subsec:strain} we investigate the impact of the entropy
strain.

\begin{figure}
\centering \psfig{file=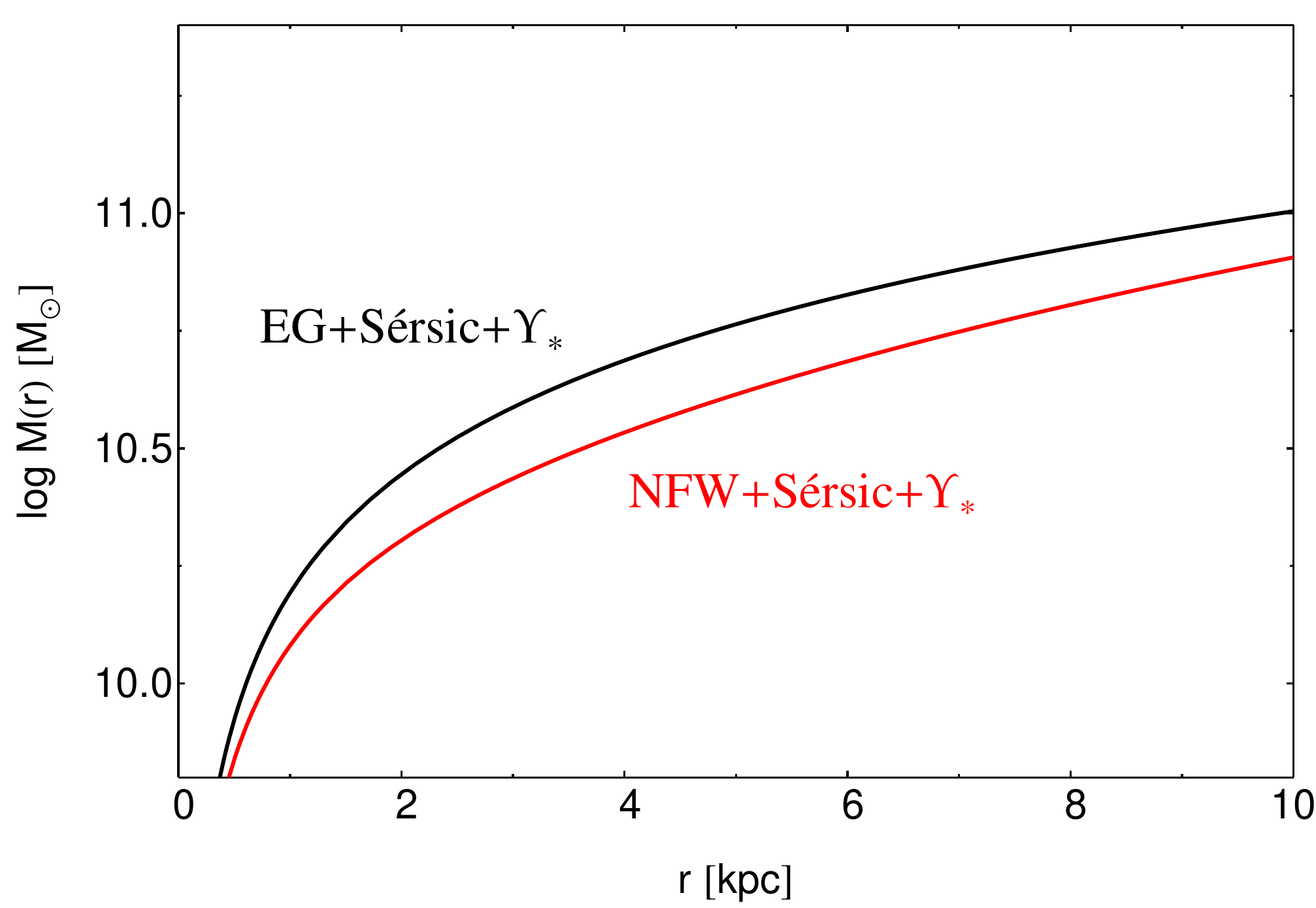, width=0.45\textwidth}
\caption{Total mass profile for EG (black line) and our NFW model
(red line). We set the parameters to the median values of the
sample: $\Re = 3.25\, \rm kpc$, $n=6.6$, the redshift to the
SPIDER average value ($z=0.08$), and $\mst = 5.4 \times 10^{10}\,
\rm \Msun$, assuming a Chabrier IMF. The NFW profile is set by
fixing \Mvir\ and \cvir\ as discussed at the beginning of
\Sec\ref{sec:Newton_and_MOND}.}\label{fig:mass}
\end{figure}

\subsection{$H_{0}$ free}\label{subsec:results_H0}

Because the value of the $H_{0}$ enters in both the distances and
the EG formula for DM, it is an interesting exercise to first
address what would happen in our modeling when it is taken as a
free parameter. For simplicity, we start adopting a universal IMF
(Chabrier or Salpeter IMF). Then, following the binning procedure
in \cite{TRN13_SPIDER_IMF} and described in
\Sec\ref{sec:Newton_and_MOND}, and adopted in particular for the
MOND models, we have created ``average'' galaxies by dividing our
sample into nine \sige-bins for which we compute median values and
$1\sigma$ scatter of all the stellar parameters relevant for our
Jeans modelling. Then, fixing the IMF, we perform a joint analysis
of the nine ``average'' galaxies and determine the best fitting
value of $H_{0}$, by minimizing a suitable "cumulative" $\chi^{2}$
function $\chi^{2} = \sum (\sigma_{\rm J} -
\sigAp)^{2}/\delta\sigma^{2}$, with $\sigma_{\rm J}$ the
theoretical $\sigma$ from the Jeans equation, and $\delta \sigma$
the error. In the minimized $\chi^{2}$, stellar masses, effective
radii and \Rap, as well as \Eq\ref{eq:veg_mdm}, depend on $H_{\rm
0}$. Finally, note that, with respect to the analysis made to
constrain the \a0\ in MOND in \cite{Tortora+14_MOND}, in this case
we are assuming that \a0\ explicitly depends on $H_{0}$, which
means that $H_{0}$ cannot vary in terms of galaxy parameters, but
has to be universal, since it is a constant of the theory.

We find that in order to match the velocity dispersions of the
galaxies in our sample with Verlinde's model, we need a
best-fitted $H_{0} = 76 \, \rm km \, s^{1} \, Mpc^{-1}$, if a
Chabrier IMF is adopted. Assuming a Salpeter IMF yields $H_{0}=
138 \, \rm km \, s^{1} \, Mpc^{-1}$. Following the discussions in
\Sec\ref{subsec:caveats}, if we assume that a) the present-day
Hubble parameter has to be consistent with predictions from
cosmological probes (e.g., Supernovae Ia or OHD observations) and
b) \Eq\ref{eq:veg_mdm} is valid, then, on average, a Chabrier-like
IMF would work for our sample of galaxies (see
\Sec\ref{subsec:caveats}). For the range of velocity dispersions
of our sample, this result is consistent with stellar populations
and independent estimates, as we will show in the next section.

\subsection{IMF free}\label{subsec:results_IMF}

In the rest of this section we adopt a value of the present day
Hubble parameter of $H_{\rm 0} = 75 \, \rm km s^{-1} Mpc^{-1}$,
consistent with local Universe measurements (e.g.,
\citealt{Riess+98_SN}; \citealt{Komatsu+11_WMAP7};
\citealt{Percival+10_BAO}; \citealt{Jimenez+03_OHD};
\citealt{Zhang+14_OHD}). It is the same used for all the results
based on DM models, and is consistent with \a0\ adopted for the
MOND models.

\Fig\ref{fig:mass} shows a typical mass profile for EG and our
reference DM-based model (i.e. the NFW+baryons), for an example
mock galaxy, setting the S\'ersic parameters and stellar mass to
the median values of the datasample, and assuming a Chabrier IMF.
The total mass from EG gravity (i.e. the sum of "apparent" DM and
stellar mass) is a factor $\sim 1.4$ larger than the total mass
derived from the DM-based NFW model, if we consider the mass
profile within \Re. To match the two profiles requires a \Yst\
value in EG smaller of a factor $\sim 1.5$ with respect to a
standard DM-based model, or equivalently a \Yst\ value in the
reference DM-based model of a factor $\sim 1.5$ larger.

To quantify the ability of EG to fit the data, we discuss the
mismatch parameter, defined as $\dimf\ \equiv \Yst/\Yst_{\rm
Chab}$, where $\Yst^{\rm Chab}$ is the \ML\ obtained by fitting
colors with stellar population models having a Chabrier IMF. In
\Fig\ref{fig:dIMF} we present \dimf\ as a function of effective
velocity dispersion \sige. We find that for EG the values of
\dimf\ are positively correlated with \sige, obtaining lower
values at the lowest-\sige\ (i.e. $\dimf \sim 0.7$) and higher
values (i.e. $\dimf \sim 1.6$) in the galaxies with the highest
\sige; on average $\dimf = 0.94$.

Below, we discuss the results shown in \Fig\ref{fig:dIMF} in more
details, contrasting the values of \dimf\ for EG, with the values
found for DM-based, MOND and stellar populations models, and we
investigate the impact of various assumptions.

\begin{itemize}

\item {\bf Panel (a).} In panel (a), the gray shaded region shows the
range of the results assuming a wide set of DM-based models,
presented in \Sec\ref{subsec:DM_models}. These results are
bracketed by the Burkert profile, producing at fixed \sige\ the
largest values of \dimf, and the AC-NFW model, which produces the
smallest values of \Yst\ and \dimf. The red line is the result for
our reference NFW profile (\citealt{TRN13_SPIDER_IMF}). On
average, the \dimf\ from EGs are $\sim 1.3$ times smaller than the
values for the reference NFW profile\footnote{Note that here the
factor $\sim 1.3$ is determined solving the Jeans equation and is
relative to the aperture of the SDSS fiber. Instead, the $\sim
1.5$ factor discussed at the beginning of this section is obtained
finding the best match between the two mass profiles in
\Fig\ref{fig:mass}, within \Re\ ($> \Rap$).}. EG models are
consistent with DM-based models adopting an AC-NFW.

\item {\bf Panel (b).} The EG results are plotted against the results
from MOND in panel (b). The cyan region is bracketed by the
results for the two standard MOND interpolating functions adopted.
The blue line adopts the interpolation function determined in
\cite{McGaugh+16} using the rotation curves of the most up-to-date
sample of spiral galaxies. The inferred values of \dimf\ from EGs
are $\sim 1.4$ times smaller that what predicted by MOND.

These results with respect to DM-based and MOND models are
consistent with what is found for the bulge components (but not
for the disks) in the sample of spiral galaxies in
\cite{Lelli+17_Verlinde}. Even larger with respect to typical
\Yst\ values are found for dwarf spheroidals
(\citealt{Diez-Tejedor+16_EG_dSph}). This is expected if we look
at the typical baryonic accelerations found in the three different
types of galactic objects. Spiral galaxies studied by
\cite{Lelli+17_Verlinde} and ETGs studied in the present paper are
characterized by similar baryonic accelerations: spirals span the
range $\sim 10^{-11.5}-10^{-9} \, \rm m \, s^{-2}$, while for ETGs
the typical accelerations spans $\sim 10^{-11}-10^{-8.5} \, \rm m
\, s^{-2}$. This results in similar values for the fitted \Yst.
While in the deep MOND regime experienced by dwarf spheroidals in
\cite{Diez-Tejedor+16_EG_dSph}, with low accelerations ($\lsim
10^{-12}\, \rm m s^{-2}$), the excess mass is larger. In this case
MOND requires values of \Yst\ that are $\sim 2.5$ times greater
than those from Verlinde's EG.

\item {\bf Panel (c).} In panel (c) of \Fig\ref{fig:dIMF} we present the two best
results from gravity-sensitive features determined from stacked
SPIDER spectra (\citealt{LaBarbera+13_SPIDERVIII_IMF}), assuming a
two-slope IMF and adopting two SSP models with the same IMF but
different ages and metallicities (solid orange line) and two SSP
models with free ages and metallicities, including, as further
free parameters, the abundances of calcium, sodium and titanium
(dashed orange line). The EG results agree with these estimates,
which are independent from either DM or gravity arguments, because
they are purely derived from the galaxy spectra and from stellar
physics.

\begin{figure*}
\centering \psfig{file=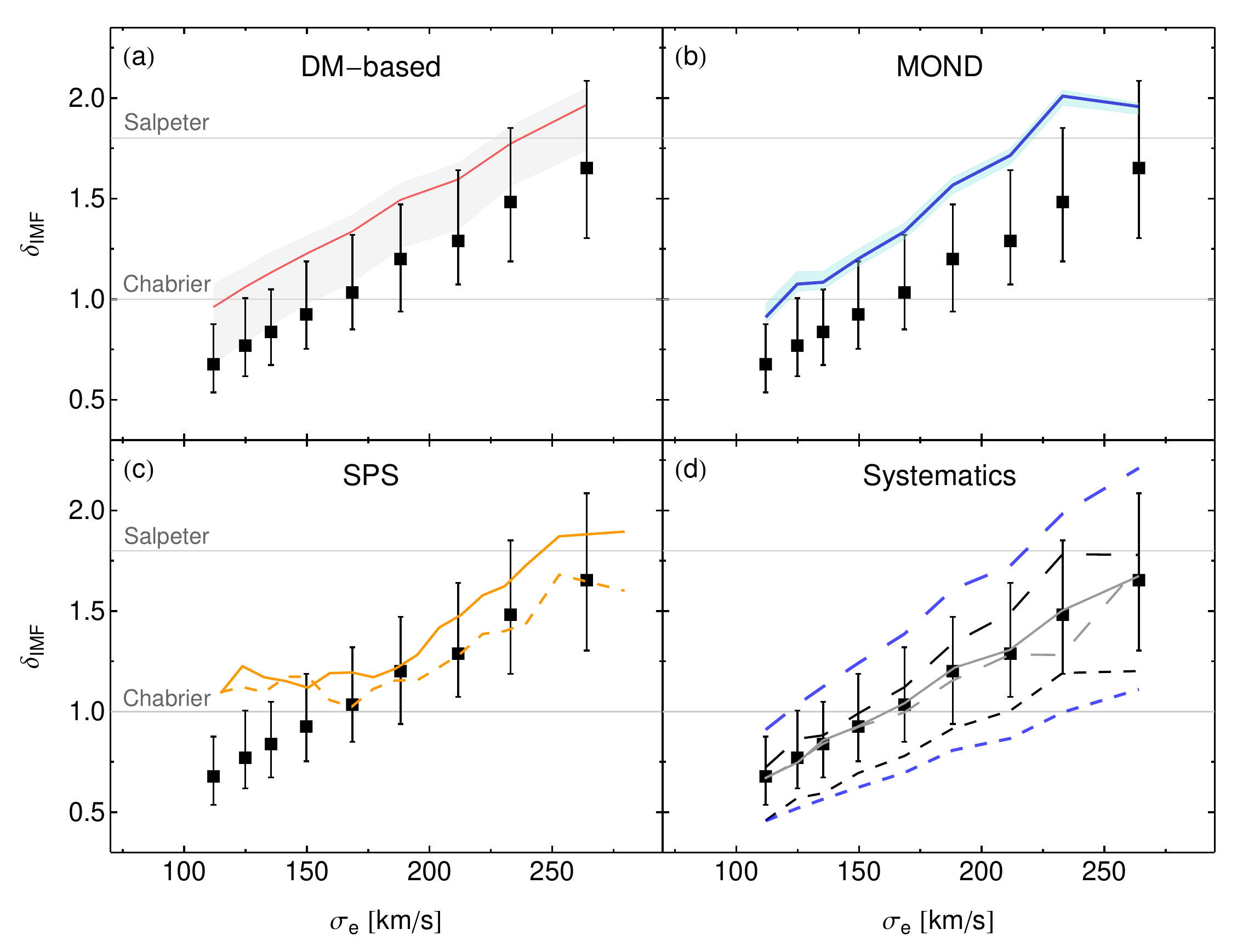, width=0.8\textwidth} \caption{IMF
mismatch parameter, \dimf, for the SPIDER ETGs, as a function of
effective velocity dispersion, \sige. Horizontal lines show the
reference values for Chabrier (bottom) and \citet{Salpeter55}
(top) IMF. Black squares and error bars are median and 25--75th
percentile trends for EG results. Comparison with other models
(DM- and MOND-based) and some systematics are shown in the
different panels. {\it Panel (a).} Red solid line plots the
medians for the fiducial DM-based model (i.e. NFW+S\'ersic).
Shaded gray region encompasses most of the DM-based models
discussed in \citet{Tortora+14_DMslope}: in particular the region
is bracketed by the Burkert and the AC-NFW models. {\it Panel
(b).} The cyan region is bracketed by the results obtained using
the two first MOND interpolating function adopted in this paper.
The blue line plots the results using the interpolating function
from \citet{McGaugh+16}. {\it Panel (c).} The orange lines are
from the analysis of gravity-sensitive features in SPIDER spectra
from \citet{LaBarbera+13_SPIDERVIII_IMF}. {\it Panel (d).} For the
EG results some systematics are investigated. Long- and
short-dashed lines are the medians when highly tangential ($\beta
= -1$) and radial ($\beta = 1$) orbits are considered,
respectively. Solid and dashed gray lines are for the roundest
objects with $q > 0.6$ and for field isolated galaxies,
respectively. Short- and long-dashed blue lines are for EG with
$H_{\rm 0} = 50$ and $100\, \rm km \, s^{-1} \,
Mpc^{-1}$.}\label{fig:dIMF}
\end{figure*}

\item {\bf Panel (d).} The effects of potential sources of systematics are shown in
panel (d). First, a possible source of systematics in the EG
results is the assumption of isotropic stellar orbits. Thus, we
have considered two extreme values of (radially constant)
anisotropy parameter $\beta$ in the Jeans equations
(\citealt{Tortora+09, SPIDER-VI, TLBN16_IMF_dwarfs}): a tangential
anisotropy, $\beta = -1$, and a radial anisotropy, $\beta = 1$,
shown in \Fig\ref{fig:dIMF} as long- and short-dashed black lines,
respectively. For tangential (radial) anisotropy larger (smaller)
\dimf\ by a factor $\sim 1.1$ ($\sim 1.3$) are found with respect
to the fiducial isotropic case. Therefore, only if strong radial
orbits in ETGs are assumed, EG does not match the results found
using DM or MOND and from gravity-sensitive features, producing
very low \dimf, unphysical at low-\sige, if compared with
predictions from synthetic models. However, detailed dynamical
modelling in the ETG central regions and simulations find
anisotropies to be fairly mild in general, typically in the range
$-0.2 \le \beta \le +0.3$ (\citealt{Gerhard+01};
\citealt{Dekel+05}; \citealt{Cappellari+07_SAURONX};
\citealt{Xu+17_Illustris}). If similar anisotropies would be found
within the EG framework, then, the impact on the inferred \dimf\
values would be negligible.

We have investigated how the assumption of spherical symmetry and
isolation of the galaxies affect our results and whether the
sample of galaxies are, on average, sufficient round and far from
companions in order to be considered a proper test-bench for EG
(see also \Sec\ref{subsec:caveats}). We first restricted the
analysis to SPIDER galaxies with K-band axial ratio $q > 0.6$ (see
gray solid line in \Fig\ref{fig:dIMF}), in order to limit to the
rounder galaxies. This subsample consists of 2847 galaxies. The
overall results are practically unchanged. Similarly, we found a
negligible impact if we limit ourselves to the 1230 field galaxies
(gray dashed line in \Fig\ref{fig:dIMF}). Consequently, the
results are unchanged if we combine these two constraints,
limiting the analysis to a sample of 807 isolated and rather round
galaxies. As for EG, this small impact of the elliptical shape and
of the environment was already verified in our previous analysis
(e.g. \citealt{SPIDER-VI}; \citealt{TRN13_SPIDER_IMF}).

In \Sec\ref{subsec:results_H0} we have determined the best-fitted
$H_{0}$ in EG for a fixed Chabrier and Salpeter IMF. Next we study
the impact of different values of $H_{0}$ on the inferred values
of \dimf. Although unrealistic, we will adopt two extreme values
$H_{\rm 0} = 50$ and $100 \, \rm km \, s^{-1} \, Mpc^{-1}$. The
results are shown in \Fig\ref{fig:dIMF} as short- (long-)dashed
blue lines, respectively, with smaller (larger) \dimf\ by a factor
$\sim 1.5$ ($\sim 1.3$) with respect to the fiducial case. $H_{0}$
values of $\sim 50\, \rm km \, s^{1} \, Mpc^{-1}$ would be
problematic for Verlinde's EG model, yielding very low stellar
\ML\ values, which particularly at low-$\sigma$ would be at odds
with predictions from spectral synthesis and independent
literature (e.g. \citealt{LaBarbera+13_SPIDERVIII_IMF}). We also
analyzed how these $H_{\rm 0}$ values impact MOND results, by
updating the adopted \a0\ and distances with the new $H_{\rm 0}$
values. We find similar changes in \dimf, which will leave
unaltered the relative discrepancy between the \dimf\ values from
EG and MOND.

\end{itemize}

In first approximation, we notice that if we would account for
small residual gas content, then the values of \Yst\ decrease for
all the results discussed in this section, potentially leaving
unaffected the difference between the different models adopted.
However, the analysis of gas contribution could be more
complicated than this simple picture, since it will follow a
different distribution than stars, potentially impacting the total
mass in EG and other models in a different way.

\subsection{Entropy strain}\label{subsec:strain}

Following \cite{Verlinde16} and all the subsequent literature, we
adopted the equality in \Eq\ref{eq:fundamentalrelation}. As we
discussed in \Sec\ref{subsec:caveats}, this is only an assumption.
There is no particular motivation to make this assumption.

In \Sec\ref{subsec:results_H0}, since the acceleration scale is a
constant in the EG proposition, we performed a joint analysis of
our galaxies, considering \a0\ to be universal. However, the
Hubble parameter in EG, and thus \a0, might be also thought to
enclose the information about the elastic medium deformation if
the principal strain does not take its maximal value. We will
investigate the case when the equality is not valid, introducing a
further dimensionless parameter $\nu_{el}$ in the Verlinde's
model, by converting \Eq\ref{eq:fundamentalrelation} in
\begin{equation}
\int_0^r \varepsilon_{\rm DM}^2 (r') A(r')dr' = \nu_{\rm
el}(r)^{2} \, V_{M_{\rm b}} (r).\label{eq:eq_nuel}
\end{equation}
The parameter $\nu_{el}$ probes the Verlinde's hypothesis of
maximum response to the entropy displacement, determining how the
elastic medium is responding to baryonic matter. If we consider
the inequality in the Verlinde's formula, then to a fixed amount
of baryons will correspond a smaller $\epsilon_{\rm DM}$ and
$M_{\rm DM}$.

We adopt $H_{\rm 0} = 75 \, \rm km \, s^{-1} \, Mpc^{-1}$ and
assume a Chabrier IMF. Looked at its face value, $\nu_{\rm el}$
acts similarly to \dimf, although it has a very different meaning,
being related to the intrinsic properties of the medium, more than
to the gas condensation properties and IMF settling. Thus,
velocity dispersions for all the SPIDER galaxies can be reproduced
if the parameter $\nu_{el}$ varies with \sige, determining a
different change of entropy displacement in terms of the baryon
mass. In fact, $\epsilon_{\rm DM}$ takes its maximal value at
lower masses/\sige, and is smaller, for larger $\nu_{el}$ in
\Eq\ref{eq:eq_nuel}, when larger masses/\sige\ galaxies are
considered.

Finally, we assume that the IMF is non universal, as found in
\Sec\ref{subsec:results_IMF}, and we infer the $\nu_{\rm el}$
which matches the typical DM-based, MOND and stellar population
results in the literature. The comparison with the results from
gravity sensitive features is good, suggesting that the effective
model of Verlinde, and in particular the assumption about equality
in \Eq\ref{eq:fundamentalrelation} might be warranted. To recover
the results from our reference DM-based model or MOND a value of
$\nu_{\rm el} \sim 1.15-1.20$ is found.

\section{Conclusions}\label{sec:conclusions}

We have studied the Emergent gravity framework from
\cite{Verlinde16}, using the central dynamics of a sample of local
massive field and round ETGs. ETGs represent an excellent test
bench of EG, because they approximate nearly spherical systems and
can also be found in more isolated environments. Under a set of
clear stated assumptions (i.e. maximum entropy strain), we show
that EG can reproduce the observed kinematics in the central
regions of ETGs, predicting stellar mass-to-light ratios, \Yst ,
similar to what is found in spiral galaxies (\citealt{Hees+17_EG};
\citealt{Lelli+17_Verlinde}). For EG, to match the central
velocity dispersion in ETGs, \Yst\ needs to be non-universal and
increasing in values from below a Chabrier to a Salpeter IMF, with
increasing stellar velocity dispersion \sige, and comparable with
stellar population studies (see \Sec\ref{sec:results} for
details). However EG produces lower values of \Yst, if compared
with independent frameworks (on average $0.94 \times \Yst_{\rm
Chab}$, and $0.7 \times \Yst_{\rm Chab}$ in the lowest-\sige\
systems). This is similar to results found recently by
\cite{Lelli+17_Verlinde} for the bulge components in a sample of
spiral galaxies. Those authors conclude that EG can be
qualitatively consistent with rotation velocity curves and the
radial acceleration relation in spiral galaxies, only if we
decrease the values of \Yst. Although these values are lower than
MOND predictions and our reference NFW+S\'ersic model, the
agreement with the \Yst\ derived from gravity-sensitive features
in SPIDER spectra or adiabatically contracted DM halo models is
quite good (\citealt{LaBarbera+13_SPIDERVIII_IMF}).

Thus, the main conclusion of this paper is that EG, DM-based
models in a Newtonian framework and MOND do reproduce the central
dynamics of ETGs and none of them can be excluded or favored.

However, in EG, observations can only put a lower bound on the
apparent DM ($M_{\rm DM}$) and acceleration (\a0). Following
\cite{Verlinde16} and the recent literature on the subject, we
assumed that the entropy strain $\varepsilon_{\rm DM}$ takes its
maximal value, a hypothesis which can be also incorrect. If we
consider the inequality in the Verlinde's formula
(\Eq\ref{eq:fundamentalrelation}), then to a given baryonic mass
of the galaxy, a lower amount of apparent DM is predicted. If we
assume that IMF and $H_{\rm 0}$ are given, then the entropy strain
$\epsilon_{\rm DM}$ has to be be maximal for galaxies with $\sigma
\sim 100 \rm \, km \, s^{-1}$ and smaller for more massive
galaxies. Moreover, if we assume that the strain is not maximal,
e.g. $\sim 1.2$ times smaller, then the central dynamics in ETGs
can only be reproduced with a higher stellar mass-to-light ratio.

More detailed analysis are needed to study the entropy strain, to
better understand the properties of the medium and the reaction to
matter displacement. We plan to further investigate the radial
mass density gradients, which can help to discriminate between EG
and other frameworks. In addition, an alternative probe is
provided by ETG strong gravitational lenses, in particular
Einstein rings which are known to have round potentials, which
through the measure of the arc radius and of the central dynamics
of the lens, allow to determine the lens mass, providing more
stringent constraints on the mass profiles and the mass density
gradients (e.g., \citealt{Barnabe+09, Barnabe+11};
\citealt{Treu+10}). It would also be interesting to study samples
of galaxies with extended (in radius) kinematical datasets, such
as Planetary Nebulae and Globular Clusters. With such galaxies, it
is possible to probe the mass distribution in the external
regions, where DM effects should be dominant (e.g.,
\citealt{Coccato+09_PNS};
\citealt{Napolitano+09_PNS,Napolitano+11_PNS};
\citealt{Pota+13_SLUGGS, Pota+15_SLUGGS}; \citealt{Alabi+16}). We
plan to test the EG formalism with these datasets, probing the
apparent mass in \Eq\ref{eq:veg_mdm} beyond the effective radius,
where the uncertainties in the stellar mass are less relevant and
the gas would contribute more to EG.

\section*{Acknowledgments}
We thank the anonymous referee for stimulating comments, which
helped us to improve the manuscript. We thank R.~H.~Sanders and
F.~La Barbera for the fruitful discussions. CT and LVEK are
supported through an NWO-VICI grant (project number 639.043.308).


\bibliographystyle{mn2e}   




\end{document}